\documentstyle[a4,12pt,epsf]{article}

\def\gtrsim{{\raise0.5ex\hbox{$>$}\hskip-0.80em\lower0.5ex\hbox{$\sim$}
}}

\begin{document}


\begin{center}
{\Large\bf{A Microscopic Energy- and Density-Dependent Effective
Interaction
and its Test by Nucleus-Nucleus Scattering}}
\footnote{Supported   by the DFG (Mu 705/3,
Graduiertenkolleg) and contract F-39 11 FMSC-PAA/NSF-94-158 (Poland)} 
\end{center}

\vspace{0.5cm}

\centerline{G. Bartnitzky$^a$, H. Clement$^a$, P. Czerski$^{b,c}$,
H. M\"uther$^b$, F. Nuoffer$^a$, J. Siegler$^a$} 
\medskip
\centerline{$^a$Physikalisches Institut, Universit\"at T\"ubingen,
D-72076 T\"ubingen, Germany}
\centerline{$^b$Institut f\"ur Theoretische Physik,
Universit\"at T\"ubingen,}
\centerline{D-72076 T\"ubingen, Germany}
\centerline{$^c$Institute of Nuclear Physics, 30152 Krakow, Poland}
\bigskip
{\bf Abstract:} An effective nucleon-nucleon interaction calculated
in  nuclear matter from the Bonn potential has been
parametrized in terms of a local density- and energy-dependent two-body 
interaction. This allows to calculate the real part of the
nucleus-nucleus scattering potential and to test this effective
interaction over a wide region of densities ($\rho \leq
3\rho_0$) produced dynamically in scattering experiments. Comparing
our calculations with empirical potentials extracted from data on
light and heavy ion scattering by model-unrestricted analysis
methods, we find quantitative agreement with the exception of proton
scattering. The failure in this case may be traced back to the properties
of the effective interaction at low densities, for which the nuclear
matter results are not reliable. The success of the
interaction at high overlap densities confirms the empirical evidence
for a soft equation of state for cold nuclear matter. \\


\newline Nucleus-Nucleus Scattering, Nuclear Matter \\

%
%

The effective interaction between nucleons bound in a nuclear medium
is of fundamental interest for the understanding of nuclear structure
aspects as well as of astrophysical problems like neutron stars and
super nova phenomena. For the latter the behavior of the effective
interaction at high densities and low temperature is crucial. In the
laboratory this region is only accessible dynamically in
nucleus-nucleus scattering, where the real central scattering
potential is strongly affected by the density dependence of the
underlying effective nucleon-nucleon (NN) interaction.
In the experiment sensitivity to the real central potential is
obtained by the observation of refractive scattering phenomena. Since
diffractive and absorptive processes dominate the scattering process
in particular for heavy scattering systems, the
observation of nuclear rainbow scattering is mandatory for a
sufficient sensitivity to the real central potential in case of
composite particle scattering. In such a case real and imaginary
parts of the potential can be extracted reliably by model-unrestricted
analysis methods from angular distributions measured with high
precision  [1-4].

Assuming a local NN-interaction the real scattering potential is obtained by
convolution of this effective NN-interaction with the point nucleon
densities of target and projectile. Antisymmetrisation in this
double folding method is  accounted for by the exchange potential,
which we have calculated  in the finite range approximation
[5,6].
For the nucleon density distributions in projectile $(\rho_{\rm P})$ and
target $(\rho_{\rm T})$ we have used the empirical results from electron
scattering. In previous double folding calculations it has been
common to factorize the effective NN interaction $v$ for convenience
into $v(s,E,\rho) = g(s,E) \cdot f(\rho)$, where $s,E,\rho$ denote
NN-distance, energy of the NN-pair and density of the
surrounding medium, respectively. For $g(s,E)$ usually a $M3Y$-type
effective interaction [7], based on the Reid-Elliot or the Paris
potential (and derived actually for a specific system, $^{16}$O,
only), has been chosen. The energy dependence then originates
exclusively from the treatment of the exchange potential. 
For $f(\rho)$ a
purely phenomenological ansatz has been made with
parameters fitted to describe scattering data for light and heavy ions. In
that  kind of double folding analyses it has been shown
[1-5,8] that
at large overlap densities $\rho > \rho_0$, where $\rho_0 = 0.17$
fm$^{-3}$ is the saturation density of nuclear matter, the density
dependence of $v$ has to be weak in order to be compatible with
scattering data, whereas for $\rho < \rho_0$ the density dependence
has to be quite strong [2]. Fig.~1 shows the results of such analyses
for $f(\rho)$ as obtained from fits to data for light ion [2,3] and heavy ion
[4]
scattering. The hatched areas give the uncertainties as estimated in
these analyses. As may be seen, the main sensitivity of heavy ion
scattering is for $\rho~\gtrsim~\rho_0$, whereas the behavior at low
densities is fixed by light ion scattering, in particular
by proton scattering [2]. We also note that for $\rho~\gtrsim~\rho_0$
our findings are compatible within the plotted error bands to the
results of Refs. [5,8] for alpha and heavy ion scattering.

In this paper we present a new, more basic approach, for which we no
longer require that the density dependence of the effective
NN-interaction factorizes  and is  adjusted purely
phenomenologically. We rather derive $v(s,E,\rho)$ without adjustment of
any parameter
from the G-matrix calculated in a nuclear matter
approach from a realistic NN-interaction.

\begin{figure}
\epsfysize=0.9\textwidth
\centerline{\epsfbox{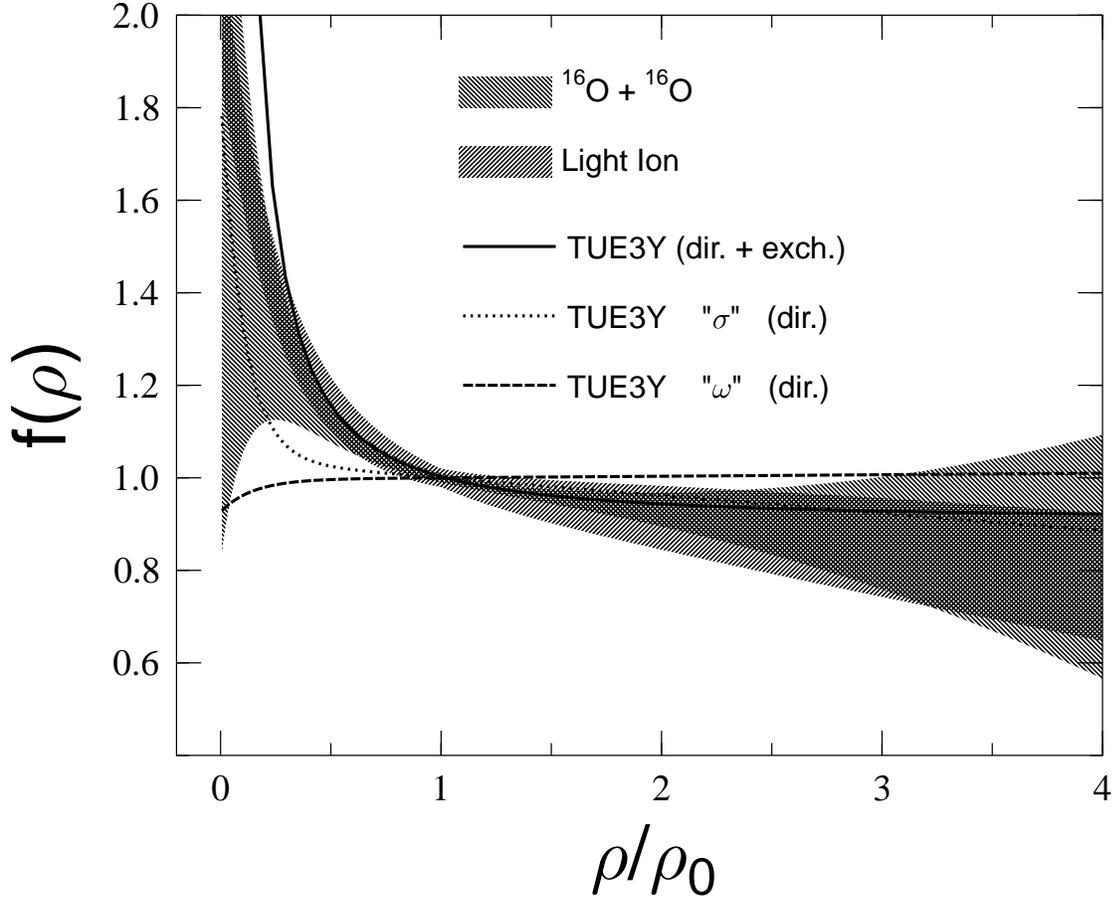}}
\caption{
Density dependence $f(\rho)$ of the effective NN-interaction
in the factorized form $v(s,E,\rho) = g(s,E) f(\rho)$. The hatched
areas  give the result of double folding analyses of light and heavy
ion scattering. The dash-dotted and dashed curves show the density
dependence of ``$\sigma$'' and ``$\omega$'' exchange in the
isoscalar-scalar part (direct term only) of our effective NN-interaction.
The solid curve gives the density dependence of the full interaction
(direct + exchange parts) for nuclear matter, i.e. without any
nuclear surface effects.}
\end{figure}

Starting point of our investigation is the $G$-matrix calculated by
solving the Bethe-Goldstone equation in nuclear matter [9,10]
\begin{equation}
G(\rho, E) = V + V \frac{ Q}{E -
Q T  Q} G(\rho, E ) \, . \label{eq:betheg}
\end{equation}
Our choice for the realistic NN potential $V$ has been the
One-Boson-Exchange potential $B$ as defined in table A.1 of
[9]. This potential is realistic in the sense that it has been
fitted to describe the NN scattering data and the properties of the
deuteron. The Bethe-Goldstone equation can be considered as an
extension of the Lippman-Schwinger equation of the reaction matrix 
for NN scattering to a system of a density $\rho$. Therefore $G$
describes the effective interaction of two nucleons in the medium,
accounting for the effects of correlations. The Pauli operator $Q$ in
(\ref{eq:betheg}), with eigenvalues 0 for two-particle states with one
of the nucleons occupying a state with momentum below the Fermi
momentum and 1 else, prevents scattering into intermediate states which
violate the Pauli principle. The Pauli operator is the source of the
density dependence of $G$. Furthermore $G$ depends on the starting
energy, $E$, which is the energy of the interacting pair of
nucleons. It is given by $E = E_{\rm kin}/A - \langle S_{\rm P} \rangle -
\langle S_{\rm T} \rangle$, where $\langle S_{\rm P} \rangle$ and $\langle S_{\rm T}
\rangle$ are the average nucleon separation energies of projectile and
target, respectively. $T$ represents the operator for the kinetic energy
and yields the energies of the intermediate particle states in the
propagator of (\ref{eq:betheg}).

In a next step we want to parametrize the effective interaction $G$ in
terms of a local two-body interaction, keeping track of the density
and energy dependence. For that purpose we consider an ansatz for the
local interaction which is of the form
\begin{equation}
 V_{\rm Local} = f(q) + f'(q)\vec{\tau_{1}} \cdot \vec{\tau_{2}} +
g(q)\vec{\sigma_{1}} \cdot \vec{\sigma_{2}} + g'(q) \vec{\sigma_{1}} \cdot
\vec{\sigma_{2}} \vec{\tau_{1}} \cdot \vec{\tau_{2}}
\; ,\label{eq:ansatz1}
\end{equation}
where the function $f(q)$ for the central scalar-isoscalar term is
parameterized as a function of the momentum transfer $q$ in terms of
two Yukawa potentials
\begin{equation}
f(q) = \sum_{i=1}^2 \frac{A_{i}}{ m_{i}^2 + q^2 } \; . \label{eq:sum}
\end{equation}
The ranges of these two Yukawa potentials, defined by the
``meson''-masses $m_{i}$ have been chosen identical to the
$M3Y$ parametrization of [7], which means $m_{1}$ = 493.3
MeV and $m_{2}$ = 789.28 MeV representing the masses of the ``$\sigma$'' and 
``$\omega$'' meson, respectively.
A corresponding ansatz has also been made for the other functions
($f'(q)$, $g(q)$, $g'(q)$) assuming the same values for the masses
$m_{i}$.  Only for the spin-isovector term ($g'$) the pion ($m_{3}$ =
139.55 MeV) has been considered explicitly, fixing the coupling
constant to the value which has also been used for the $M3Y$
parametrization.
The parameters for the coupling constants ($A_{i}$, and corresponding
ones for the other channels)
have been adjusted such that the local interaction $V_{\rm Local}$
fits the antisymmetrized matrix elements calculated for the $G$-matrix
at a given density $\rho$ and starting energy $E$, i.e. $A_i =
A_i(\rho,E)$ etc. This means that
at each energy and density, the $G$-matrix is parametrized in terms of
8 coupling constants.  The required effective NN-interaction
$v(s,E,\rho)$ is then just given by the Fourier transformation
of (\ref{eq:ansatz1}) and (\ref{eq:sum}), respectively. As a result
one finds that the energy- and momentum-dependence of 
$A_i$ in general is rather weak. Strong deviations are only observed 
at small densities and large
starting energies $E$. This special
behavior of the parameters at small densities may be interpreted as
an indication that the attempt to derive a local effective interaction
from the $G$-matrix of nuclear matter at densities far below the 
saturation density does not produce very reliable results. This may
reflect the fact that the homogeneous nuclear matter at these densities
is instable against the formation of droplets.
Hence for this density region a theoretical treatment
starting from finite nuclei is expected to be more reliable. 

In the following we compare double folding potentials calculated from
the microscopically derived interaction $v(s,E,\rho)$ to empirical
potentials obtained from model-unrestricted analyses of proton, alpha
and $^{16}$O scattering from  closed shell nuclei. These have been
chosen in order to minimize coupled channel effects as well as the
dominance of absorption phenomena in the scattering process. While
proton scattering is sensitive to low densities $\rho < \rho_0$, the
largest overlap densities up to $3\rho_0$ occur in
$\alpha$-scattering due to the large central density of the
$\alpha$-particle. Heavy ion scattering on the other hand provides
quite extended spatial  regions of density overlap in the course of
the scattering process coming thus closer to the dynamical
realization of nuclear matter at high density. $^{16}$O  + $^{16}$O
is the hitherto only heavy ion system where refractive nuclear
rainbow scattering has been observed [4,5].

\begin{figure}
\epsfysize=0.8\textwidth
\centerline{\epsfbox{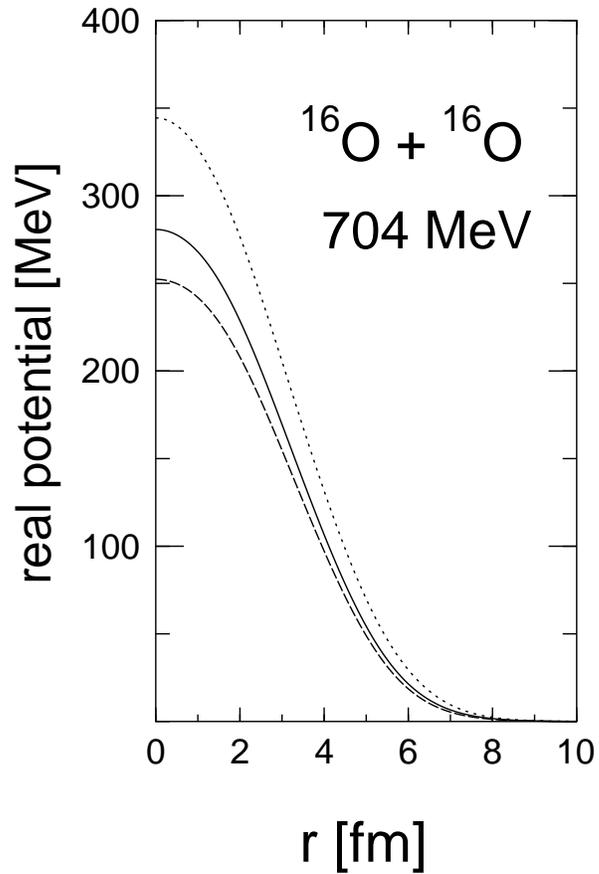}}
\caption{ 
Double folding calculations for the real central scattering
potential of $^{16}$O + $^{16}$O at $E = 704$ MeV. The solid curve
shows the result by use of the correct choice of the starting energy,
whereas for the other curves either the nucleon separation energies
(dotted) or the kinetic energy between projectile and target (dashed)
are neglected in the definition of the starting energy.}
\end{figure}

The energy dependence of the double folding potential originates from
two different sources. One is the non-locality of the Fock-exchange
terms, which yields a momentum-dependent local potential. The second
source is the explicit energy dependence of the underlying effective
NN interaction, see eq. (1). In order to emphasize the importance of
this explicit energy dependence we compare in Fig. 2 the real part of
the optical potential with the choice for the starting energy as
discussed above (solid) with results obtained, when the off-shell or
binding effects are ignored (dotted line, $E = E_{\rm kin}/A,~\langle S_{\rm P} \rangle =
\langle S_{\rm T} \rangle = 0$). Ignoring Pauli effects this would
correspond to an approach, which uses the T-matrix rather than the
G-matrix for the effective NN interaction. One observes that this
approach overestimates the real part by up to 20\%. If the kinetic
energy between projectile and target is ignored in the definition of
the starting energy, the dashed line is obtained. This choice of the
starting energy would be appropriate for calculating the binding
energies of nuclei. We see that attempts which try to derive the
nuclear equation of state from scattering experiments [5]  without accounting 
for the explicit dependence of the interaction on the starting energy
tend to overestimate the binding energy of nuclear matter as compared
to more rigorous calculations.

Fig.~3 displays a comparison of our
double folding potentials derived from the G-matrix 
to the empirical potentials
extracted from scattering data by model-unrestricted analysis
[2,4].
The hatched areas represent the  uncertainties in the
extracted potentials. We see that without adjustment of any parameter
the calculations yield already a nearly quantitative description of
the empirical potentials. Small readjustment of the potential
strength $(\lambda = 0.9 - 1.0)$ leads to full agreement with the
experimental results\footnote{We note that also the descriptions of
the corresponding experimental angular distributions are 
comparable to those obtained in the model-unrestricted
analyses.}, with the exception of the proton scattering
potentials. There the calculated potentials show a peculiar, bumpy
behavior at the surface, which is not observed experimentally and
which leads to a root mean square radius, which in case of $p
+~^{40}$Ca is larger by 30\% than the experimental value and thus far
beyond the experimental uncertainty. At a first glance this
discrepancy is very surprising, since the conventional method of
using a factorized effective interaction has led to a very
satisfactory description of real central proton scattering potentials
[2]. The failure of our more rigorous calculation can be traced back
to the very different density dependence of ``$\sigma$'' and
``$\omega$'' exchange in the effective interaction at low densities,
as is apparent from fig.~1. Since in the conventional calculations
with factorized interactions implicitly identical density
dependencies are assumed for ``$\sigma$'' and ``$\omega$'' exchange, this
problem has not been encountered there. On the other hand the failure
of the nuclear matter calculations for $\rho < \rho_0$ is not totally
unexpected as we discussed above.

\begin{figure}
\epsfysize=0.7\textwidth
\centerline{\epsfbox{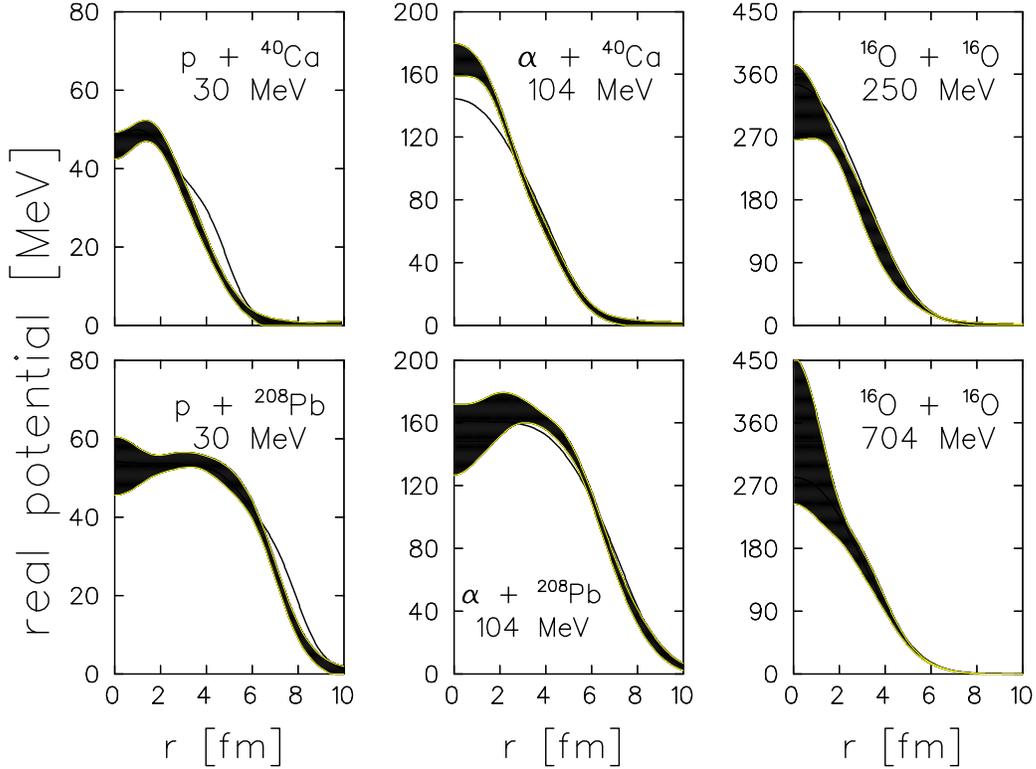}}
\caption{
Fig. 3: Real central scattering potentials for the systems $p
+~^{40}$Ca, $p +~^{208}$Pb, $\alpha +~^{40}$Ca, $\alpha +~^{208}$Pb
and $^{16}$O + $^{16}$O at $E/A = 22 - 44$ MeV. The hatched areas
represent the empirical potentials [2,4] with their
uncertainties. The results of our double folding calculations are
shown by the solid lines.}
\end{figure}

In summary we have exploited the strong sensitivity of the real
nucleus-nucleus scattering potentials to the density dependence of
the underlying effective NN-interaction to test a new theoretical
derivation of the latter, starting from nuclear matter calculations.
This inherently energy- and density-dependent interaction
$v(s,E,\rho)$ reproduces
empirical heavy and light ion scattering potentials (real part) on a
quantitative level, with the exception of proton scattering. The
failure there can be traced back to the problems of parametrizing 
a $G$-matrix for nuclear matter at densities $\rho <\rho_0$.
At high densities, $\rho >
\rho_0$, the good agreement with the results from  scattering data
confirms the weak density dependence of $v(s,E,\rho)$ which 
leads [4,9,10] to a soft equation
of state for nuclear matter with a compressibility of $K \approx 190$
MeV. We note, however, that the nuclear matter saturation properties
calculated with $v(s,E,\rho),~~E/A = -13$ MeV and $\rho_0 = 0.22$
fm$^{-3}$, are slightly off the empirical values $(-16$ MeV, 0.17
fm$^{-3}$) in resemblance of the well-known Coester-band problem
[9].

\end{document}